\documentclass[a4paper]{article}

\usepackage{icrc2013} 
\usepackage[english]{babel}

\title{Characterization of Potential U.S. Sites for the Cherenkov Telescope Array}

\shorttitle{U.S. Sites for CTA}

\authors{
R.A. Ong$^{1}$,
T. Aune$^{1}$ and
J. Hall$^{2}$,
for the CTA Consortium.
}

\afiliations{
$^1$ Department of Physics and Astronomy, University of California, 
Los Angeles, CA 90095, USA \\
$^2$ Lowell Observatory, 1400 W. Mars Hill Road, Flagstaff, AZ 86001, USA \\
\scriptsize{
}
}

\email{rene@astro.ucla.edu}

\abstract{
The Cherenkov Telescope Array (CTA) is a major ground-based observatory
proposed for gamma-ray astronomy.  CTA is envisioned to consist
of two large arrays of atmospheric Cherenkov telescopes for the study
of sources of high-energy gamma rays in the energy range of a few tens 
of GeV to beyond 100 TeV.  One array would be located in the southern
hemisphere and one in the northern hemisphere.  After a detailed
search, we have identified two potential sites in the USA for the 
northern array.  Both sites are located in northern Arizona.
Here we describe the two sites and the deployment of instrumentation 
to characterize them.  The characteristics of the sites,
in terms of their atmospheric and climatic properties,
are described. We show recent data from the automated monitoring
equipment at the sites and compare these data to a commercial simulation.
Details regarding the facilities and infrastructure required for the sites
are also presented.
}


\begin{document}
\maketitle

%Begin a section.

\section{Introduction}
The Cherenkov Telescope Array (CTA) is an international consortium of 27 countries 
and approximately 1,000 scientists working to build the next generation 
gamma-ray observatory using the atmospheric Cherenkov technique and 
improving by a factor of ten the sensitivity of 
ground-based facilities for very high energy (VHE) gamma-ray astronomy.  CTA will comprise 
two arrays, one in each hemisphere. Essential physical characteristics of the sites 
include 1 km$^2$ of relatively flat land (with $< 8$\% grade across the entire area) and 
elevation $> 1500$ m.    The standard CTA atmospheric monitoring station, the “ATMOSCOPE”, 
has been installed at each site since June, 2012.

The US team of CTA has identified two sites in Arizona as candidates for the 
northern hemisphere array.  Land in Arizona has a variety of ownership types (private, 
State, US national forest, and Native American reservation).  The proposed CTA sites 
are both on 1.6 km x 1.6 km square sections of privately owned land, which allows 
maximum ease and promptness of leasing and permitting.   
Figure 1 shows maps of the sites, illustrating the “checkerboard” pattern of land ownership 
prevalent in Arizona.  White sections are privately owned. 

 \begin{figure}[t]
  \centering
  \includegraphics[width=0.45\textwidth]{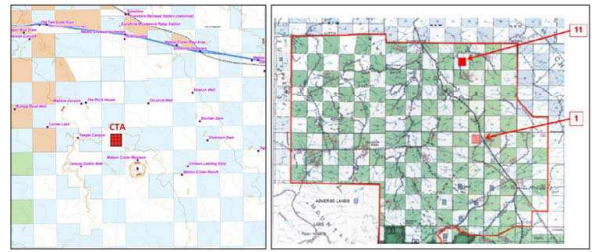}
  \caption{
Maps of the sites.  Left: Arizona East, near Meteor Crater.   The four-lane Interstate 40 is the blue line, 
6 km north of the site (red square).  Right: Arizona West, 25 km south of the small town of Seligman, AZ, on Interstate 40.  Green sections of land are the Prescott National Forest.  The US team investigated sites 1 and 11 as indicated by the red squares. Site 11 (upper square) is the proposed one for CTA.
          }
 \end{figure}

\section{The Arizona Sites}

The Arizona East site is located 3 km from Meteor Crater and 65 km from the city of Flagstaff (pop. 65,500).  
The site is almost completely flat and treeless, and it lies at an elevation of 1,677 m.   One of the US four-lane national highways, Interstate 40, is 6 km north and the main line of the Burlington Northern Santa Fe railroad 
is 1 km farther north of the proposed site.

The Arizona West site is located on the Yavapai Ranch, 25 km south of the town of Seligman, AZ (pop. 400), at an elevation of 1,670 m.  Vegetation at the site is the low juniper pine and pinyon pine characteristic of this elevation in Arizona.  This site is less flat than Arizona east but remains well
within the CTA requirements.

The sites are reasonably close to Flagstaff, requiring either 
a 50-minute (Arizona East) or a 1 hour 45 minute (Arizona West) drive to reach.  
Lowell Observatory, which 
could serve as the local managing and scientific partner for the CTA, is sited 
in Flagstaff and permanent CTA employees would likely be based there.

 \begin{figure}[t]
  \centering
  \includegraphics[width=0.45\textwidth]{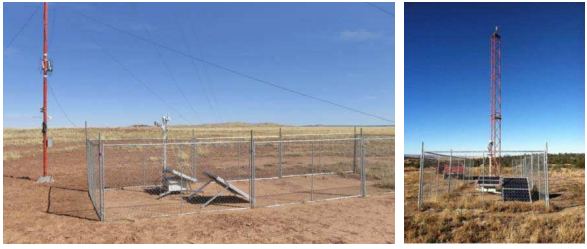}
  \caption{
Photographs of the sites.  
Left: Arizona East, with the lip of Meteor Crater in the distance.  
Right: Arizona West. The CTA ATMOSCOPEs are visible in each fenced enclosure, 
as are the 30-m (left) and 12-m (right) anemometer towers.
          }
 \end{figure}

\section{Meteorological and Atmospheric Characteristics of the Sites}

\subsection{General Characteristics}
Both sites have decades to over 100 years of meteorological records in place.  
All significant variables (temperature, humidity, wind, and natural hazards) in 
the historical record are within the CTA survivability 
requirements and within the
operating requirements for a high fraction of the time.
Data from the CTA ATMOSCOPEs also show no episodes of weather outside 
the requirements as well.

\begin{table}[ht]
\begin{center}
\begin{tabular}{|l|c|c|}
\hline Quantity & Arizona E & Arizona W \\ \hline
Ave max temp   & 34.2$^\circ$C (Jul)  & 33.3 $^\circ$C (Jul) \\ \hline
Ave min temp   & -6.8$^\circ$C (Jan) & -7.5$^\circ$C (Dec) \\ \hline
Ave precipitation & 19.3\,cm/yr  & 33.2\,cm/yr \\ \hline
Ave snowfall & 28.2\,cm/yr & 13.0\,cm/yr \\ \hline
\end{tabular}
\caption{Long-term average climate values for the proposed Arizona sites.
The duration of climatic records for the Arizona E(W) site is
118(20) years.}
\label{table_single}
\end{center}
\end{table}

\subsection{Night Sky Background}
In addition to the ATMOSCOPE measurements, we 
commissioned sky brightness measurements from the 
US National Park Service.  An example of the results
is given in Figure 3, showing the 
all-sky image for the Meteor Crater site.  This is a V-band image 
of the entire sky taken at the site in September 2012.
The light domes from nearby communities are on the horizon and  
none encroach on the CTA operational field of view 
$> 30$ degrees above the horizon.

 \begin{figure}[t]
  \centering
  \includegraphics[width=0.45\textwidth]{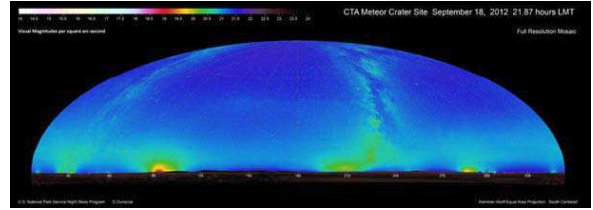}
  \caption{
All-sky panorama of the night sky brightness at Arizona East
site in Arizona, USA. At zenith, the site brightness is
21.7 magnitudes arcsec$^{-2}$, essentially free of anthropogenic
glow.  The circular arcs extended to higher elevations are due
to the Milky Way.
          }
 \end{figure}

\subsection{Studies using the CTA ATMOSCOPES}

To precisely characterize the meteorological and observational 
conditions at the CTA candidate sites, self-contained devices, 
dubbed ATMOSCOPEs, were installed at both the Meteor Crater and Yavapai Ranch
locations in northern Arizona. The ATMOSCOPEs consist of a number of instruments 
that measure wind speed, wind direction, temperature, relative humidity, 
atmospheric pressure, night sky background
light, and cloud cover. The ATMOSCOPEs were installed at both sites at the end of 
June, 2012 and have been operating nearly continuously since that time. 
Additional instrumentation for each ATMOSCOPE (the all-sky camera and Sky Quality
Monitor) were installed in September 2012.
At that same time, two mirrors of the VERITAS design were installed
at the Arizona East site to study the effects of weathering at the site.
These mirrors were placed on the tower, 8\,m
above the ground and oriented facing north and south.
Some details about the ATMOSCOPEs and the
various instruments can be found in \cite{bib:Bulik}.

During the nearly full year of ATMOSCOPE operation at the Arizona sites, 
detailed information on the
environmental conditions (wind, temperature, humidity)
 have been acquired and analyzed. 
Of particular interest are the wind speeds. It was found that 
average wind speeds have a strong
correlation with the time of day, being significantly 
higher during the day than at night. Figure 4 shows
the average wind speed versus hour of the day for both 
CTA candidate sites in Arizona. 

 \begin{figure}[t]
  \centering
  \includegraphics[width=0.52\textwidth]{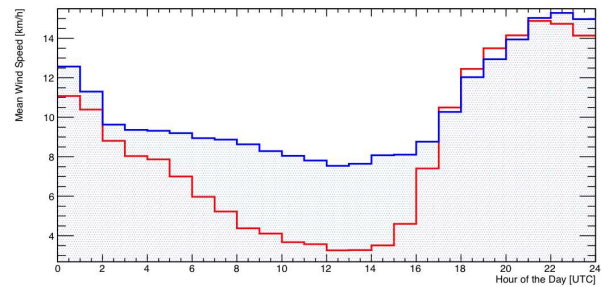}
  \caption{
Average wind speeds at the Arizona East (blue) and
Arizona West (red) sites as a function of the
hour of day, as measured by the CTA ATMOSCOPEs.
Night time corresponds approximately to UTC hours 
of 03-12.
These data come from the period between late June 2012
and early March 2013.
          }
 \end{figure}

While the ATMOSCOPEs have provided very detailed information 
since their installation, estimating the
suitability of the sites for the 10+ years of the CTA experiment's presumed 
lifetime calls for an
understanding of the longer-term trends. 
In addition to the long-term weather records mentioned
previously, retrodictions of environmental conditions at the CTA candidate sites
have been obtained from weather simulations carried out by a private company (SENES)
(see \cite{bib:Bulik}).
Correlation studies with simultaneous ATMOSCOPE and
weather simulation data have been 
performed and the results from these studies give us a
picture of the long-term weather patterns.

As an example, Figure 5 shows the correlation between the temperature
as recorded by the ATMOSCOPE and the prediction of the simulation for the
Arizona East site.  The data are essentially simultaneous (i.e. within
one hour overlap, the time resolution for the SENES study).
As can be seen, the correlation is relatively good with
a moderate amount of spread. On the other hand, Figure 6 shows the
comparison for the wind speed at the same site.  
Here it is clear that the the simulation overestimates the wind speed.
Thus a correction factor (that varies with wind speed)
is needed in order to correctly use the simulation to
estimate, for example, the fraction of time when the wind speed is
expected to be above
the CTA operational requirement.

\begin{figure}[t]
  \centering
  \includegraphics[width=0.45\textwidth]{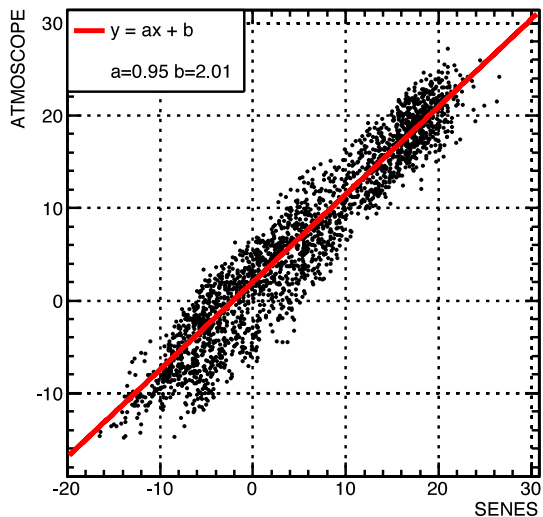}
  \caption{
Comparison of the temperature as measured by
the ATMOSCOPE and the temperature predicted by the
simulation (SENES) at the Arizona East site. 
The red line (and associated inset) gives a linear
fit to these data.
These data come from the period between late June 2012
and early March 2013.
          }
\end{figure}

\begin{figure}[t]
  \centering
  \includegraphics[width=0.45\textwidth]{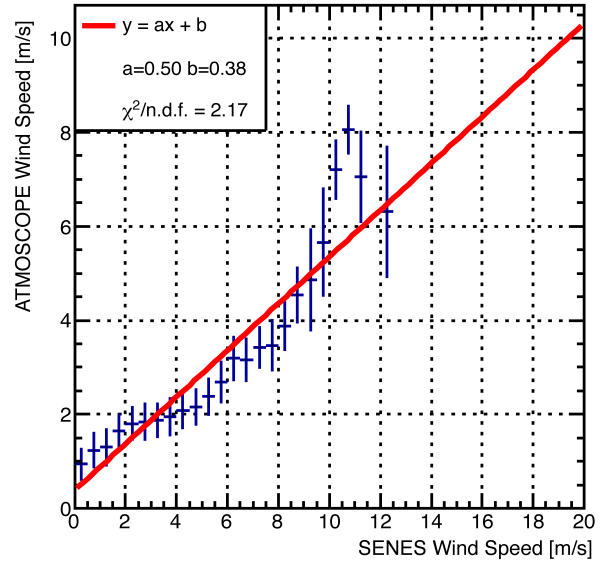}
  \caption{
Comparison of the wind speed as measured by
the ATMOSCOPE and the speed predicted by the
simulation (SENES) at the Arizona East site. 
The red line (and associated inset) gives a linear
fit to these data.
Since the simple linear fit does not describe the data very well,
a two-component linear fit is used to describe the relation
between predicted and measured wind speed.
These data come from the period between late June 2012
and early March 2013.
          }
\end{figure}

To quantify the darkness of the night sky at the candidate CTA sites, 
two devices are employed in the ATMOSCOPE \cite{bib:Gaug}. One
is the Sky Quality Monitor (SQM) which is a commercial device 
produced by Unihedron. It utilizes a
TSL237 photodiode and a temperature sensor to provide a measure of the night 
sky brightness in mag/arcsec$^{2}$. It has a claimed precision of 
$\pm 10$\% ($\pm 0.10$ mag/arcsec$^{2}$). The
spectral responsivity of photodiode is quite broad, extending from 300 nm 
to 1100 nm with the peak
responsivity at $\sim 700$ nm. This broad spectral response makes it a 
less-than-ideal instrument for measuring the night sky background relevant 
for atmospheric Cherenkov telescopes. The second
device on the ATMOSCOPE used to measure the night sky background is the Light of Night Sky (LoNS)
device. The key elements of the LoNS are a Hamamatsu 3584 PIN-photodiode and a filter wheel which
includes a V-band filter and a custom ``B'' filter made from a combination of Schott BG25 and BG39
filters - meant to emulate the response of a photomultiplier tube. The data from the LoNS is the
current measured from the photodiode (in pA). 

In order to validate the night sky background measurements and to 
try to establish a conversion
between LoNS photodiode current (in pA) and night sky background 
light (in mag/arcsec$^{2}$), the
correlation between the LoNS ``B''-filter photodiode current and the SQM measurement was
investigated. This procedure was also used for the LoNS V-band filter. 
The correlation scatter plot (LoNS ``B'' vs. SQM) and unbinned fit are shown in Figure 7. 
Based on the definition of magnitude, one expects the correlation to follow the 
function ${\rm SQM} = a \log_{10}({\rm LoNS}) +
b$ where $a=-2.5$. Due to the difference in spectral response and fields of view (SQM FWHM $\sim
20^{\circ}$, LoNS FWHM $\sim 40^{\circ}$), the theoretical correlation is not seen in either the
``B'' or V-band LoNS filters. However, the relatively tight correlation does allow for a conversion
from narrow-band LoNS current measurements to broadband measure of the night sky background in
mag/arcsec$^{2}$.  We find that the ATMOSCOPE
estimates for the night sky background levels at both sites
agree with those determined by photometric techniques and carried out by the
US National Park Service (see, e.g., Figure 3).

\begin{figure}[t]
  \centering
  \includegraphics[width=0.52\textwidth]{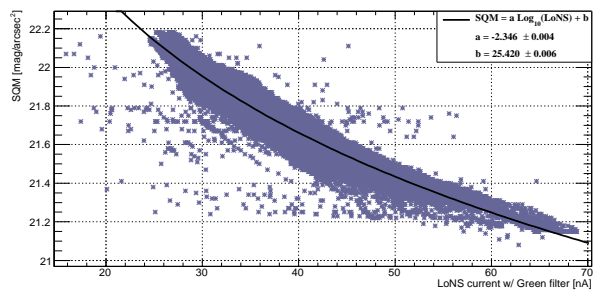}
  \caption{
Comparison of the night sky light conditions as measured by
the SQM device (Sky Quality Monitor, see text for details)
and the LoNS device (Light of Night Sky, see text for details),
both incorporated into the CTA ATMOSCOPE.
The fit to the data, shown by the black curve, is
described in the text.
          }
\end{figure}

\section{Facilities and Infrastructure}

Both sites offer easy access and convenient infrastructure development.  
A modern, four-lane highway runs to within 7 km of AZ East and 25 km of AZ West, 
with either paved or good dirt roads coming to within a few kilometers of the sites.  
Electric substations operated by Arizona Public Service (APS)  are nearby 
(within about 8 km for the East site). 
Wells are available at both sites for water, and internet connectivity up to 
1 Gbps is available with high-speed microwave connections using either direct or 1-hop 
lines of sight to mountaintop towers near Flagstaff.

Nearby support facilities are excellent.  Flagstaff has a modern airport 
with a 2,700 m runway and instrument landing system, and both Phoenix Sky Harbor 
International Airport and Las Vegas McCarran International Airport are 3-4 hour 
drives from the sites.  Flagstaff Medical Center is a 267-bed, 
Level 1 Trauma Center facility equipped to handle any medical emergency 
and it is served by an air transport service that can reach the hospital 
from Arizona East in 15 minutes and from Arizona West in 35 minutes.  
Most permanent CTA personnel would likely be based in Flagstaff.

\section{Conclusions}
After a search of possible locations in the United States for the northern site
of the Cherenkov Telescope Array, we have identify two sites in
northern Arizona that fully satisfy all CTA requirements.
Here we provide some of the general characteristics for the sites,
including typical meteorological conditions, night sky background levels,
and available facilities and infrastructure.
Results from the CTA ATMOSCOPE monitoring stations are shown, including
initial comparisons between weather conditions measured by the ATMOSCOPE
and those predicted by a commercial simulation.
These comparisons give us some confidence in the ability of the simulation
to estimate the weather conditions over a longer period of time.

\vspace*{0.5cm}
\footnotesize{{\bf Acknowledgment:}{
We gratefully acknowledge support from the agencies and organizations 
listed in this page: http://www.cta-observatory.org/?q=node/22}}.
This work was carried out, in part, with 
support from the US National Science Foundation, from
the University of California and
from the Economic Collaborative of Northern Arizona (ECoNA).
We are grateful for the assistance of Judy Prosser and Fred Ruskin.

\end{document}